\documentclass{iaus}
\usepackage{graphics}

  \checkfont{eurm10}
  \iffontfound
    \IfFileExists{upmath.sty}
      {\typeout{^^JFound AMS Euler Roman fonts on the system,
                   using the 'upmath' package.^^J}%
       \usepackage{upmath}}
      {\typeout{^^JFound AMS Euler Roman fonts on the system, but you
                   dont seem to have the}%
       \typeout{'upmath' package installed. iaus.cls can take advantage
                 of these fonts,^^Jif you use 'upmath' package.^^J}%
      }
  \else
  \fi

  \checkfont{msam10}
  \iffontfound
    \IfFileExists{amssymb.sty}
      {\typeout{^^JFound AMS Symbol fonts on the system, using the
                'amssymb' package.^^J}%
       \usepackage{amssymb}%

      }{}
  \fi

  \IfFileExists{amsbsy.sty}
    {\typeout{^^JFound the 'amsbsy' package on the system, using it.^^J}%
     \usepackage{amsbsy}}
    {\providecommand\boldsymbol[1]{\mbox{\boldmath $##1$}}}

\newsavebox{\astrutbox}
\sbox{\astrutbox}{\rule[-5pt]{0pt}{20pt}}

\title[The Interplay among Black Holes, Stars and ISM in Galactic 
       Nuclei]{Large-- and Small--Scale Feedback in the Starburst 
Centers of Galaxies}

\author[D. Calzetti {\it \/}]%
{Daniela Calzetti$^1$%
}

\affiliation{$^1$Space Telescope Science Institute, Baltimore, MD 21218, 
USA email: calzetti@stsci.edu}

\pubyear{2004}
\volume{222}
\pagerange{1--8}
\date{?? and in revised form ??}
\jname{The Interplay among Black Holes, Stars and ISM \\in Galactic Nuclei}
\editors{Th. Storchi Bergmann, L.C. Ho \& H.R. Schmitt, eds.}
\begin{document}

\maketitle

\begin{abstract}
The interplay between stellar populations and gas in local starburst
galaxies is analyzed using images from the Hubble Space Telescope to
map the ages of the young stellar components and to isolate the
contribution of shocks on spatial scales ranging from a few tens of pc
to $\sim$1~kpc. The shocked gas represents a small fraction of the
total ionized gas in these objects, yet it can have profound effects
on the long--term evolution of the starburst, which may include the
triggering of new star formation.
\end{abstract}

\firstsection 
\section{Introduction}

The evolution of galaxies and of the intergalactic/intracluster medium
(IGM/ICM) are closely connected, via the continuous interchange
between the ISM and the IGM/ICM. The heating and metal enrichment of
the IGM/ICM, and the enrichment of galaxies' ISM with pristine gas via
infall from the IGM are manifestations of this close
connection. Evidence for gas outflows from galaxies has been found at
redshift as high as $\sim$3 (\cite{pettini00}), and may have 
contributed to the early pollution of the IGM (\cite{ellison00}).

The driving mechanism for the energy and metal pollution of the
IGM/ICM is the feedback from star formation. In the basic scenario,
the combined effects of ejecta and energy release from supernovae and
massive stars in regions of intense star formation form superbubbles
that sweep up the surrounding medium and produce cavities containing
hot, shocked gas (\cite[Weaver et al. 1977]{weaver77}). If the bubble
has enough energy to expand to a size of a few times the scale height
of the ambient medium, the swept-up shell may break out and the wind
internal to the bubble may expand into the IGM (\cite[Chevalier \&
Clegg 1985]{chevalier85}).

Many theoretical and observational studies have investigated the
mechanical feedback from star formation and its efficiency in
transfering energy and material to regions far removed from the site
of star formation (\cite[e.g., De Young \& Heckman 1994]{deyoung94};
\cite[Suchkov et al. 1994]{suchkov94}; \cite[Mac Low \& Ferrara
1999]{maclow99}; \cite[Strickland \& Stevens 2000]{strickland00};
\cite[Heckman, Armus \& Miley 1990]{heckman90}; \cite[Dahlem
1997]{dahlem97}; \cite[Ferrara \& Tolstoy 2000]{ferrara00};
\cite[Cecil, Bland-Hawthorn \& Veilleux 2002]{cecil02}), as well as in
triggering renewed star formation (\cite[e.g., Wada \& Norman
2001]{wada01}). Yet many questions remain to be fully answered. Among
these, progressing from small to large scales: the impact of the
spatial distribution, history, and duration of the star--forming event
(\cite[Clarke \& Oey 2002]{clarke02}); the coupling of the feedback
from star formation with the surrounding ISM; the dependence of the
feedback efficiency on the conditions of the ISM (e.g. porosity;
\cite[Clarke \& Oey 2002]{clarke02}); and the role of the global
galaxy parameters (mass, metallicity, environment; \cite[Dahlem
1997]{dahlem97}; \cite[Strickland et al. 2004]{strickland04}).  A
review of some of these issues is given by D. Strickland (2004, these
proceedings).

One of the complicating factors in tackling the various facets of
feedback is the vast range of physical scales that needs to be
probed. While the relevant scales for charting star formation are
those of the stars and star clusters (a few pc), a study of the
structure and conditions of the ISM needs to sample scales of tens to
hundreds of pc. Moving on to larger scales, superwinds are best
investigated on kpc scales, while probing the influence of global
parameters involves scales of tens of kpc (galactic sizes) to Mpc
(interactions). 

Even for the closest galaxies, the small--to--intermediate scales
relevant for the interaction feedback--ISM are accessible only via
high angular resolution observations (e.g., with the Hubble Space
Telescope), as 10~pc=0.4$^{\prime\prime}$ for a galaxy at 5~Mpc
distance. The study reviewed here involves four local starburst
galaxies observed with the Wide Field Planetary Camera~2 on the
HST. Because of their intense-to-extreme star formation rates,
starburst galaxies are the sites where stellar feedback can have its
most dramatic influence on the structure and evolution of the ISM (and
surrounding IGM; \cite{heckman90}), and are, therefore, optimal
laboratories to investigate mechanical feedback. The four galaxies in
the present study cover a range in luminosity, metallicity, star
formation rate, and environment, but they are all closer than 5~Mpc
(Table~\ref{tab1}). This investigation attempts at addressing the
issue of the coupling feedback--ISM, and its relation to the duration
of the star forming event. The problem is tackled from two fronts:
stellar populations and ISM conditions. The ages of the young
($\lesssim$300~Myr) stellar clusters are used to set a minimum value
to the duration of the current starburst event, while measurements of
the shocked gas provide a constraint on the fraction of starburst
mechanical output that is recovered in the ISM within a small distance
($\lesssim$1~kpc) of the site of star formation.

\begin{table}
  \begin{center}
  \begin{tabular}{lrrrrrrr}
      Galaxy  & $D$   & $d$ & $M_B$ &   $(O/H)$ & $SFR(H\alpha)$ &     
$    L(H\alpha)_{nph}/L(H\alpha)$ &    $    L(H\alpha)^{pred}_{nph}/L(H\alpha)_{nph}$\\
      Name    & (Mpc)   & (kpc) & &           & (M$_{\odot}$~yr$^{-1}$) & & \\[3pt]
        NGC3077&3.85&1.4&$-17.5$&8.9&0.076&0.043&0.51--1.40\\
        NGC4214&2.94&2.0&$-17.2$&8.2&0.089&0.041&0.29--1.24\\
        NGC5236&4.5&0.75&$-20.3$&9.2&0.308&0.030&1.27--2.90\\
        NGC5253&4.0&1.5&$-17.5$&8.2&0.270&0.032&0.38--1.19\\
  \end{tabular}
  \caption{Properties of the observed galaxies. Columns are as
  follows; $D$: distance; $d$: physical size of the region observed;
  $M_B$: absolute B magnitude; $(O/H)$: oxygen abundance,
  12$+$log(O/H); $SFR(H\alpha)$: star formation rate derived from the
  extinction corrected H$\alpha$ luminosity;
  $L(H\alpha)_{nph}/L(H\alpha)$: fraction of the H$\alpha$ luminosity
  associated with shocks; $L(H\alpha)^{pred}_{nph}/L(H\alpha)_{nph}$:
  ratio of the predicted-to-observed H$\alpha$ luminosity associated to
  shocks. NGC3077 is in the M81 group, in close interaction with M81
  itself; NGC4214 is an isolated galaxy, while NGC5236 and NGC5253
  form a loose pair.}
  \label{tab1}
  \end{center}
\end{table}

\section{Charting the Young Stellar Populations in Starbursts}\label{sec:starpops}

The star cluster populations and the spatial distributions of their
ages within the starburst sites were investigated for three of the
galaxies listed in Table~\ref{tab1}, NGC3077, NGC5236, and NGC5253
(\cite{harris01}, \cite{harris04}). For NGC5253, the diffuse starburst
population had previously been age-dated (\cite{calzetti97},
\cite{tremonti01}), which, combined with the more recent study of its
cluster population, yields a complete picture of the recent star
formation history in this galaxy.

\begin{figure}
\centering 
\resizebox{6.5cm}{!}{\includegraphics{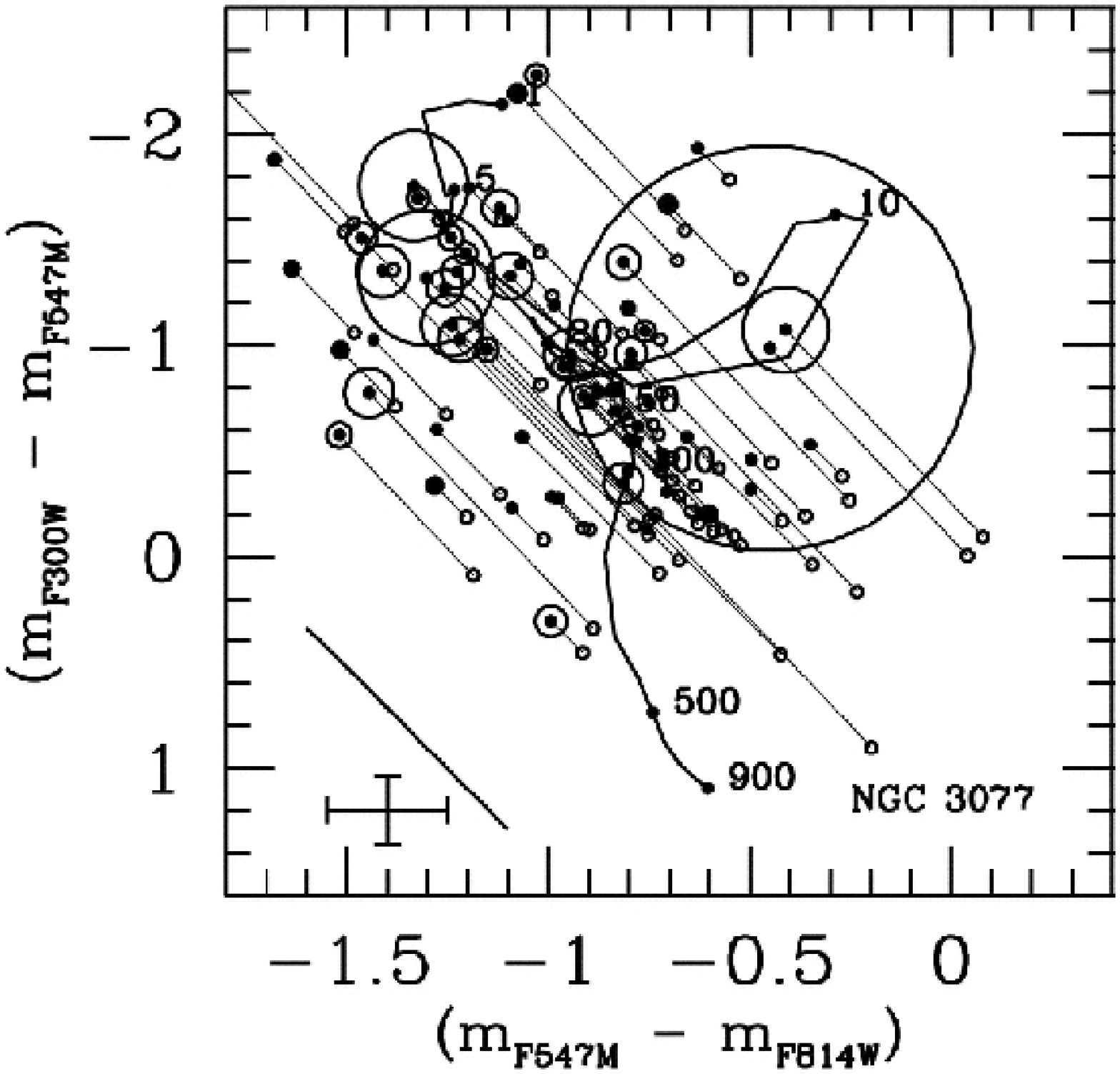}}
\resizebox{6.5cm}{!}{\includegraphics{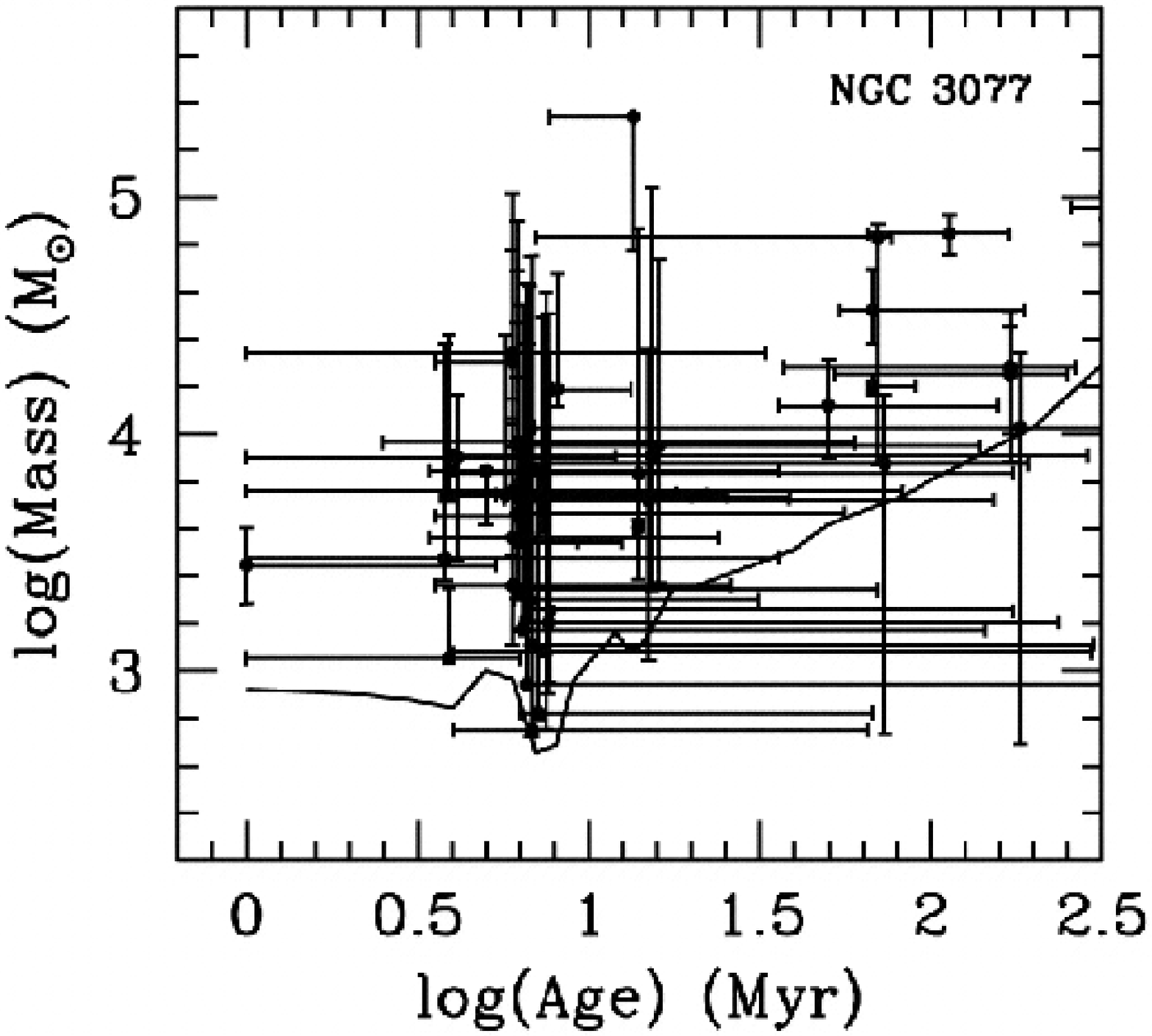}}
\caption[]{({\bf Left}). The color--color diagram, UV$-$V versus V$-$I
(in HST magnitudes), of the stellar clusters in NGC3077. The small
open points are the observed data, while the filled points are the
reddening--corrected data, with the direction of the reddening vector
shown as a diagonal line at the bottom--left of the plot. The two sets
of points are connected by lines. The sizes of the circles concentric
to the filled points are proportional to the total V--band flux of
each clusters. The colors of an instantaneous burst model from
Starburst99 (\cite{leitherer99}) are shown for the age range
1--900~Myr as the continuous line across the diagram. ({\bf
Right}). The age--mass plot of the stellar clusters in NGC3077. Each
data point is shown with its error bar. The curve at the bottom
represents the 90\% completeness limit.  Both figures are from
\cite{harris04}.}\label{fig1}
\end{figure}

Images of the galaxies in the HST/WFPC2 medium and broad filters UV,
V, I, and in the H$\alpha$ and H$\beta$ line emission provide enough
angular resolution to identify clusters, and enough color information
to age-date them. In particular, two age--sensitive diagnostics, the
color--color diagram UV$-$V~versus~V$-$I and the equivalent width of
H$\alpha$, were combined to constrain the most likely age for each
stellar cluster and, assuming a Salpeter IMF, to assign a mass
(Figure~\ref{fig1}; \cite{harris01},
\cite{harris04}). H$\alpha$/H$\beta$ ratio maps were used to correct
photometry and fluxes for the effects of dust reddening
(\cite{calzetti94}). The cluster ages were then re-mapped onto the
galaxy images to derive the history of the cluster formation in each
starburst.

The two dwarf galaxies, NGC3077 and NGC5253, show time--extended star
formation, spanning the last 100--300~Myr, and a very `chaotic'
spatial distribution of the cluster ages. This similarity between the
two galaxies is counterbalanced by striking differences. In NGC3077,
there are star clusters as old as 300~Myr, while in NGC5253 the oldest
clusters unambiguously detected in the HST images do not seem to be
significantly older than $\sim$20~Myr (\cite{harris04}). Indeed, the
time--extended star formation history for NGC5253 is inferred not from
its clusters, but from its diffuse stellar population, whose colors
are consistent with constant star formation over the past
$\sim$100-200~Myr (\cite{calzetti97}). The dearth of `old' star
clusters in NGC5253 could be caused by rapid cluster dissolution
timescales in the center of the galaxy; by rescaling the Milky Way
model of \cite{kim99}, \cite{tremonti01} and \cite{harris04} derive
dissolution timescales in the range 16--50~Myr for NGC5253. These
short timescales are driven by the high central velocity dispersion
measured in the galaxy (\cite{caldwell89}). A smaller velocity
dispersion may be present in NGC3077, as inferred from molecular cloud
velocities (\cite{walter02}), implying longer survival times for its
clusters.

Contrary to the dwarfs, the distribution of the stellar clusters in
the center of the giant spiral NGC5236 shows a high level of
organization, possibly a reflection of the different dynamical
environment (presence of a bar) in the massive galaxy. The UV--bright
clusters are distributed along a half--ringlet surrounding the
optically--bright nucleus. The distribution of the ages also shows a
high level of correlation in the star formation, which appears
propagating in the S--N direction and inside--out in the ringlet
(\cite[Harris et al. 2001]{harris01}, \cite{puxley97}). There is a
clear peak in the number of clusters in the narrow age range 5--7~Myr,
with a handful of older clusters (up to $\sim$50~Myr). Although the
diffuse population is generally consistent with constant star
formation over the past $\sim$1~Gyr, the sharp cut--off of the cluster
ages around 7~Myr could still be explained with either a recent 
starburst or very rapid cluster dissolution timescales
(\cite{harris01}).

Independently of whether the current starburst in the grand design
spiral is confirmed to be younger than $\sim$10~Myr, a clear mechanism
for the triggering and feeding of the starburst, i.e. gas infall along
the main bar, is present in NGC5236 (\cite{gallais91}). No such
obvious mechanism is present in the two dwarf galaxies. Their
starbursts may have been triggered by past interactions with their
companions (M81 for NGC3077, and NGC5236 for NGC5253), but these
interactions occurred a few hundred Myr ago, and are unlikely to be
the direct cause of the starbursts observed today.

\section{Charting the Shock--Ionized Gas}\label{sec:shocks}

\begin{figure}
\centering 
\resizebox{6.5cm}{!}{\includegraphics{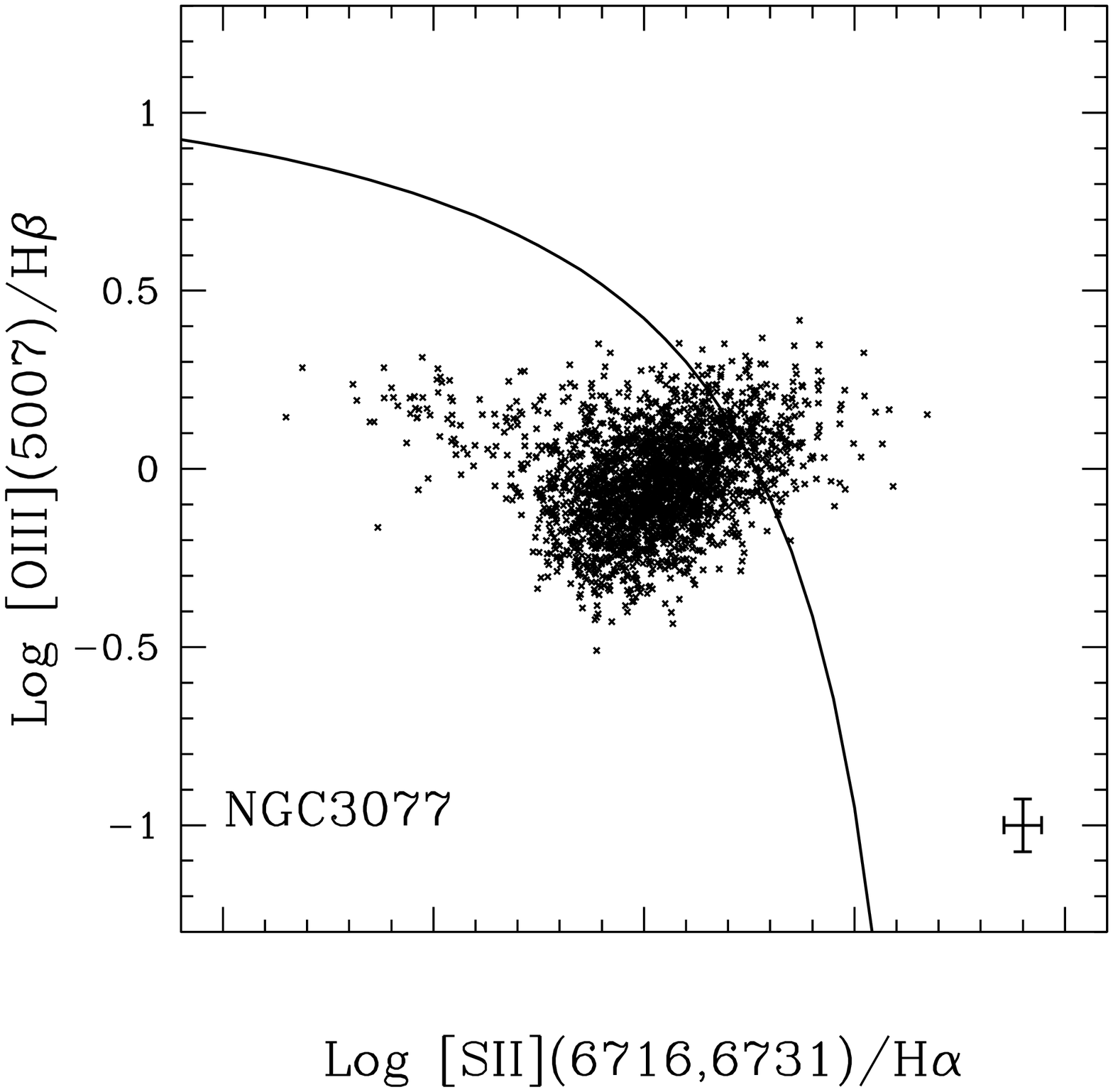}}
\resizebox{6.5cm}{!}{\includegraphics{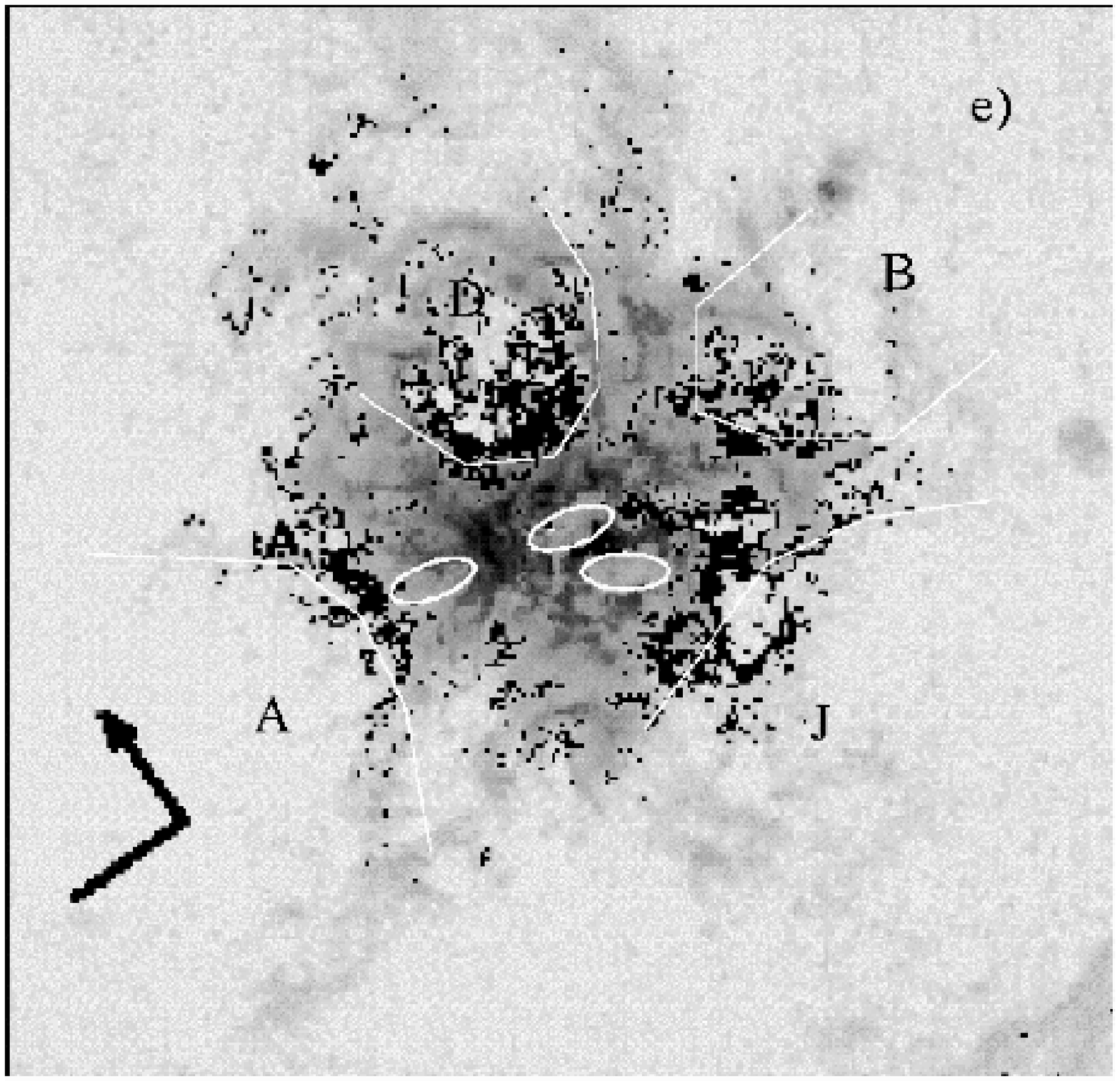}}
\caption[]{({\bf Left}). The
log([OIII]/H$\beta$)--versus--log([SII]/H$\alpha$) diagnostic diagram
for NGC3077. Each data point represents a bin
0.3$^{\prime\prime}\times$0.3$^{\prime\prime}$ in the WFPC2 images, or
about 5.6$\times$5.6~pc$^2$ at the distance of the galaxy
(Table~\ref{tab1}). Typical 1~$\sigma$ uncertainties are shown at the
bottom right of the plot. The `maximum starburst' line of
\cite{kewley01} is shown as a continuous line. This line is used to
separate areas dominated by photoionizing processes (below-left of the
line) from non--photoionizing processes. ({\bf Right}). The bins
identified as dominated by non-photoionization (shocks) in the diagram
at left are reported in black on a gray-scale H$\alpha$ image of
NGC3077. North is indicated by the arrow. The position of H$\alpha$
shells is shown by white poligons, with their naming convention
(\cite{martin98}), while the positions of the CO emission
(\cite[Walter et al. 2002]{walter02}) are shown as white
ellipses. Both figures are from \cite{calzetti04}.}\label{fig2}
\end{figure}

The stellar population study of these galaxies was complemented with 
the investigation of the conditions of the ISM within and
around the four starbursts. The standard emission line diagnostic
diagram log([OIII]/H$\beta$)--log([SII]/H$\alpha$) was used to
discriminate photoionized from non--photoionized gas
(\cite{baldwin81}). HST/WFPC2 images were obtained with narrow--band
filters in correspondence of the forbidden lines [OIII](5007~\AA) and
[SII](6726~\AA), and of the hydrogen recombination lines H$\alpha$ and
H$\beta$. The diagnostic diagram was constructed by dividing each
image in a grid of bins 3--6~pc in size, measuring line fluxes in each
bin 5~$\sigma$ above the noise level, and deriving line ratios from
spatially corresponding bins. Figure~\ref{fig2} (left~panel) gives an
example of the resulting diagrams (\cite{calzetti04}).

The `maximum starburst line' defined by \cite{kewley01} was used as
the separating line between photoionized gas (to the bottom--left of
the line) from non--photoionized gas (to the top--right of the line,
Figure~\ref{fig2}). This can be considered a conservative 
definition of non--photoionized gas, yet it 
yielded clear identifications in all four galaxies. The non--photoionized gas 
in these galaxies shows two basic morphologies: enclosed bubbles and 
shells/filaments (Figure~\ref{fig2}, right panel), and is
likely due to shock ionization (\cite{martin97}). Indeed, the presence of
star-formation-induced shocks is supported by observations of diffuse
X--ray emission in the centers of all four galaxies. The
shock--ionized component is responsible for a small fraction, about
3\%--4\%, of the total H$\alpha$ luminosity (Table~\ref{tab1},
column~7), but is spread over a significant fraction of the area covered by
the H$\alpha$ emission, about 15\%--25\%.

The availability of multi--color (UV--to--I) and H$\alpha$ images
allows us to place contraints on the characteristics of the ionizing
stellar population, and, once coupled with models, on the expected
amount of mechanical energy coming out of the starburst.  The
H$\alpha$ luminosity predicted from the mechanical starburst output
can then be compared with the {\it observed} H$\alpha$ luminosity
associated with shocks. The last column of Table~\ref{tab1} shows the
predicted-to-observed H$\alpha$ luminosities from shocks, with the
range marking starbursts of different durations, from 10~Myr
(left-hand-side figure) to 100~Myr (right-hand side figure). In
this column, a figure smaller then 1 means that the starburst does not
produce enough mechanical energy to account for the observed H$\alpha$
luminosity of shocks, while a figure larger than 1 means the starburst
can support the observed shock luminosity.

For the three dwarfs, between 70\% and 100\% of the mechanical energy is 
deposited within the immediate surroundings of the starburst, roughly within 
the inner 0.2--1~kpc. Furthermore, starbursts with durations shorter than 
$\sim$30~Myr do not produce enough mechanical energy to account for 
the observed H$\alpha$ luminosity of shocks. It is worth reminding that 
we adopted a conservative criterion for identifying shocked gas in the 
galaxies, thus the limitation on the minimum duration of the starbursts is 
quite stringent. In addition, the result is independent of whether the parent 
galaxy is isolated or in interaction with other galaxies. 
Since the vast majority of the starburst mechanical output is recovered 
in the immediate surroundings of the starburst itself, it is not unlikely 
that the shocks can trigger renewed star formation. This inference could 
help explain the long star formation timescales required by the high 
luminosity of the 
H$\alpha$ emission associated with shocks, and the even longer timescales 
derived from the stellar population studies. 

For the giant spiral, only between 35\% and 75\% of the available
mechanical output is recovered in the surrounding ISM, for a variety
of starburst conditions. Thus, there are no restrictions on the
current starburst coming from the characteristics of the shocked
gas. In addition, the non--photoionized gas is located in two main
bubbles, and there is a striking absence of extended features (shells
or filaments).  This suggests a confined starburst.

\section{Conclusions}\label{sec:concl}

The characteristics of the stellar populations and ISM in the
starbursts of four nearby galaxies suggest a strong coupling between
the starbursts and the surrounding ISM. This is particularly true for
the three dwarf galaxies in the sample. Here the amount of observed
H$\alpha$ luminosity associated with shocks corresponds to about
70\%--100\% of the total mechanical output from the starbursts, and
places a lower limit of about 30~Myr to the duration of the starbursts
(i.e., these cannot be `bursts'). Such large fractions of mechanical
energy available in the immediate surroundings of starbursts may
suggest triggered star formation as a mechanism to explain the long
duration timescales inferred from stellar population studies
($\approx$100--300~Myr), in the absence of direct, recent triggers.

\begin{acknowledgments}
The author would like to thank the Local Organizing Committee, and in
particular Thaisa Storchi-Bergmann, for their hospitality.  This work
was supported by NASA Long--Term Space Astrophysics grant NAG-9173 and
by NASA HST grant GO-9144.
\end{acknowledgments}

\end{document}